\documentclass[preprint] {article}
\usepackage[utf8]{inputenc} 
\usepackage{amsmath}
\usepackage{amsfonts}
\usepackage{amssymb}
\usepackage{geometry}
\allowdisplaybreaks
\newcommand {\bp}{\begin{pmatrix}}
\newcommand {\ep}{\end{pmatrix}}
\newcommand{\be}{\begin{equation}} \newcommand{\ee}{\end{equation}}
\newcommand{\bea}{\begin{eqnarray}}\newcommand{\eea}{\end{eqnarray}}

\usepackage{geometry} 
\begin{document}
\title{Integrable nonlocal vector nonlinear
Schrödinger equation with self-induced parity-time-symmetric
Potential
}
\author{Debdeep Sinha\footnote{{\bf email:}  debdeepsinha.rs@visva-bharati.ac.in} 
\ and Pijush K. Ghosh \footnote {{\bf email:}
pijushkanti.ghosh@visva-bharati.ac.in}}
\date{Department of Physics, Siksha-Bhavana, \\ 
Visva-Bharati University, \\
Santiniketan, PIN 731 235, India.}
\maketitle
\begin{abstract}
A two component nonlocal vector nonlinear Schrödinger equation (VNLSE) is
considered with a self-induced $ {\cal PT}$ symmetric potential.
It is shown that the system possess a Lax pair and an infinite number
of conserved quantities and hence integrable. Some of the conserved quantities like number
operator, Hamiltonian etc. are found to be real-valued, in spite of these charges
being non-hermitian. The soliton solution for the same equation is
obtained through the method of inverse scattering transformation
and the condition of reduction from nonlocal to local case is also
mentioned. 
An inhomogeneous version of this VNLSE with space -time modulated
nonlinear interaction term is also considered
and a mapping of this Eq. with standard VNLSE through  similarity transformation
 is used to generate its solutions.

\end{abstract}

\tableofcontents
\vspace{0.3in}


\section{Introduction}
The advent of $\cal{PT}$-symmetric quantum mechanics \cite{aa} 
opens up the possibility of having real spectra corresponding 
to non-Hermitian Hamiltonians. This stimulates a great deal of 
research work concerning non-Hermitian Hamiltonians  
\cite{ali,znojil,piju,fring,longhi,fabio,bikash}.
 In the realm of non-relativistic quantum mechanics, 
the $\cal{PT}$-symmetric Hamiltonian yields real spectra
as far as $\cal{PT}$-symmetry remains spontaneously unbroken,
 on the other hand,
 if the $\cal{PT}$-symmetry is broken spontaneously
the spectrum 
is not entirely real.  
 The paraxial 
equation of diffraction, being similar in structure with the 
Schrödinger equation, strongly suggests that optics might
 be a testing ground of $\cal{PT}$-symmetric systems 
\cite{kg}.This possibility stimulated a great deal of research works 
\cite{r1,MU1, r10} in $\cal{PT}$-symmetric optical 
systems with the experimental observation of  the
$\cal{PT}$-symmetric phase transition between 
broken and unbroken $\cal{PT}$-symmetric phases \cite{cer}. Also,
the possibility of observing various interesting physical phenomena 
such as power oscillation, double refraction and secondary emission
in ${\cal PT}$-symmetric periodic potentials is discussed in the
literature\cite{kg, MU1}.

The governing equation 
in nonlinear optics is the NLSE describing wave 
propagation in nonlinear media \cite{ac}. 
NLSE is an integrable nonlinear equation admitting soliton
solution. Apart from describing wave propagation
in non-linear media \cite{ac}, NLSE finds a wide class of applications in 
many physical systems such as
in  Bose-Einstein
condensation (BEC) \cite{r2}, in plasma physics \cite{r3},
in gravity waves \cite{r4},
in $\alpha$-helix protein dynamics\cite{r5} etc..
Thus in this regards the study of various 
$\cal{PT}$-symmetric  versions of NLSE may 
be very interesting.

\par 
An integrable $\cal{PT}$-symmetric nonlocal nonlinear 
Schrödinger equation (NLSE), arising due to a nonlocal 
reduction in the Lax-pair formulation,  
was introduced in Ref. \cite{abm} for which exact soliton 
solutions were obtained through  inverse scattering
 methods. In this equation the Schrödinger field and 
its parity transformed complex conjugate are treated 
as two independent fields. In Ref. \cite{aks} it was 
shown that this nonlocal NLSE admits dark as well 
as bright soliton solutions in case of attractive 
potentials and several periodic solutions for
 the same equation was obtained in Ref. 
\cite{akas}. 
A discrete version of the nonlocal NLSE is considered
in Ref. \cite{akmu} which is shown to be integrable and 
possess a Hamiltonian structure. A one soliton
solution is also obtained for this model \cite{akmu}. In Ref. \cite{li} 
a chain of nonsingular localized wave solution was obtained 
for nonlocal NLSE via a 
Darboux transformation. Apart from that 
various kind of 
multicomponent soliton interaction was also considered \cite{li}. 
In Ref. \cite{lym} it was shown that the nonlocal NLSE and its discrete
version are gauge equivalent to the corresponding version of Heisenberg
equation and modified Heisenberg like equation respectively. 

A generalized
version of nonlocal NLSE in an external potential
with a space-time modulated coefficient of the 
nonlinear interaction term and/or gain-loss terms
 is considered in Ref. \cite{dp}. A mapping of this Eq. 
to the standard NLSE through
similarity transformation is used to find exact
solution of inhomogeneous and/or non-autonomous
nonlocal NLSE. Further a $(d+1)$-dimensional 
generalization of nonlocal NLSE without the 
external potential is considered and is shown to
possess all the symmetry of the Schrödinger 
group. The conserved  charges corresponding space-time symmetries turned out
 to be real-valued in spite of these charges being non-Hermitian.
Finally, the dynamics of different moments are studied with a
 time evolution of the
“pseudo-width” of the wave packet in case
 the system possess a O(2,1) conformal symmetry.

\par 
A possible generalization of NLSE is the
vector NLSE that arises in physical situations whenever
two wave-trains in nonlinear medium travel nearly with the same 
group velocity \cite{Byj, Brg}. Moreover, vector NLSE 
manifest itself, in physical systems  such as in wave-guides
and in optical fiber where the involved electric field has more
than one components \cite{Besg}. The vector NLSE
also arises in the study of multi-component
BEC, for example,
the dynamics of various moments in the context
of multicomponent BEC is considered in \cite{pkg1}  
Both NLSE and  VLSE admits integrable 
discretization that are used in numerical study of the 
corresponding continuous case as well as to model
some discrete physical systems \cite{abcd,abcdt}.  
In Ref \cite{akas}, a two component generalization of the nonlocal
 NLSE was introduced and some of its exact soliton 
solutions both having vanishing as well as 
non-vanishing boundary conditions at infinity were written down.  

 In this paper we consider the multicomponent
 generalization of nonlocal NSLE and study its integrability.
 In particular, we use the Lax pair formulation as well as the method of 
inverse scattering transformation to obtain soliton 
solution of the nonlocal VNLSE. We obtain
an infinite number of conserved quantities and relate some
of them with the known symmetries of the system. 
Some of the conserved charges are shown to be real valued,
although those are  non-Hermitian in nature. An inhomogeneous 
non autonomous version of the 
nonlocal VNLSE  is also considered and a mapping of this Eq. with normal VNLSE through 
similarity transformation is employed to generate its solutions. A few physically interesting
examples are presented.
 
\par

The plan of this paper is as follows. We first formulate 
the scattering problem by splitting the nonlinear evolution
 Eq. into two parts, one of which is an eigenvalue 
problem with constant eigenvalue and other one
 gives the time evolution of the corresponding fields.
 In the next section we discuss the direct scattering 
problem which is relevant in constructing the 
eigenfunctions of the eigenvalue problem and 
discuss their linear independence, symmetry and 
construct the proper eigenvalues. In the following
 section we shall take the other linear Eq., i.e, the 
time evolution Eq., to find the time evolution of 
the involved fields. In the section $V$ we construct  
the inverse scattering problem as a Riemann-
Hilbert (RH) problem which reduces 
a set of linear-integral Eqs. to a set of linear algebraic
 Eqs. that can be solved to find the solution of the 
nonlinear Eq. In the very next section we construct 
the one soliton solution for the two component 
nonlocal NLSE. The 
conserved quantities are discussed in section $VII$. Section $VIII$ deals with
an inhomogeneous  version of this nonlocal VNLSE  with 
space-time modulated nonlinear interaction term and construct
its solution using the similarity transformation method.
 In the last section we make a 
summary and discuss the results.

\section{Introduction to the model and Formulation of the scattering problem}

We have considered the following two component vector 
nonlinear Schrödinger (VNLSE) equation:
\bea
i{\bf Q}_t={\bf Q}_{xx}+2{\bf Q}{\bf Q}^{{\bf P}}{\bf Q}
\label{1e}
\eea
\noindent where  ${\bf{Q}}=(q_1(x,t), q_2(x,t))$ is a two component
row matrix with $q_{1}(x,t), q_{2}(x,t)$ being two complex fields having
the real arguments $x$ and $t$ and ${\bf P}$ denotes a 
operation that implies hermitian conjugation plus parity transformation
of the spatial argument. In particular,
$${\bf Q}^{{\bf P}}= \bp
{q^{*}_1(-x,t)}\\
{ q^{*}_2(-x,t)} \ep .$$
\noindent In contrast to 
the normal two component vector NLSE, here in the potential term, the  
argument of the complex field is $-x$ instead of $x$. This introduces 
a nonlocal nature of the potential. 
The nonlocal non-linearity is commonplace
in nature. For example it appears in case of diffusion 
of charge carriers, atom
 or molecules in atomic vapors \cite{AC, DT} as well as 
in the study of BEC with a 
long range interaction, one of such example is
 the BEC with magnetic dipole-dipole forces 
 \cite{GR1}. The BEC of chromium is investigated 
in Ref. \cite{GR2} and the optical 
spatial soliton in highly nonlocal medium is 
observed in Ref. \cite{ncry}. The 
non-linearity of the type as indicated in Ref.
 \cite{abm} is supposed to be realized in case of  
 coupled wave guide or in an infinite array of 
wave guides system \cite{aks}. It may be recalled that 
a few exact solutions of Eq. (\ref{1e}) were written down in
Ref \cite{akas}. The main emphasis of this paper is to
study the integrable properties of the system governing Eq. (\ref{1e})
by employing standard techniques like Lax pair formulation and inverse
scattering transformation and find soliton solutions.

The set of equations (\ref{1e}) can be obtained from the following
Lagrangian density:
\bea
{\cal L}=\frac{i}{2}\left[{\bf Q}{\bf Q}_t^{{\bf P}}-{\bf Q}_t{\bf Q}^{{\bf P}}\right]
+{\bf Q}_x{\bf Q}_x^{{\bf P}}+({\bf Q}{\bf Q}^{{\bf P}})^2.
\label{lag}
\eea
\noindent The Lagrangian density ${\cal{L}}$ and the corresponding action
is invariant under space-time translation and a boost. The corresponding
conserved quantities are non-hermitian. It will be shown in Sec. $VII$ that
some of the conserved charges are  real-valued in spite of these quantities
 being non-hermitian. The Lagrangian density ${\cal{L}}$ admits a global
$SU(2)$ symmetry and the corresponding conserved quantities have the
expressions:
\bea
N^a={\bf Q}\sigma^a{\bf Q}^p
\label{N}
\eea
\noindent 
where $\sigma^a$ with $a=1, 2, 3$ are the Pauli matrices.
It is interesting to note that $Ns$ are neither hermitian nor a
semi-positive definite quantity. Later we shall show 
that $Ns$ are real-valued.

Eq. (\ref{1e}) admits a Lax pair formulation in case of 
which the nonlinear equation yields a pair of linear 
equations 

\bea
v_{x}&=&
\bp
-ik {\bf I}_{1\times1}& {\bf Q}\\
 {\bf R} & ik  {\bf I}_{2\times2} \\
\label{sc}
\ep
\label{2e}
v\\
v_{t}&=&
\bp
2ik^2+i {\bf Q} {\bf R}& -2k {\bf Q}-i {\bf Q}_{x} \\
-2k {\bf R}+i {\bf R}_{x} &  -2ik^2 {\bf I_{2\times2}}-i {\bf R} {\bf Q} \\
\ep v\nonumber\\
\label{3e}
\eea
one of which is a eigenvalue problem with a constant eigenvalue k
and other one gives the time evolution of the corresponding fields. 
Here ${\bf I}_{n \times n}$ denotes $ n \times n$ square matrix.
In this case $v(x,t)$ is a three component column vector and 
the fields $\{q_1(x,t),q_2(x,t)\}$ and $\{r_1(x,t),r_2(x,t)\}$
 vanish rapidly as $|x|$ $\rightarrow$ $ \infty$. The compatibility
 condition $v_{xt}$ $=$ $v_{tx}$ between eqs. (\ref{2e}) and (\ref{3e}) 
give rise to 
eq. (\ref{1e}) under the symmetry reduction:
\be
\bf{R}=-\bf{Q}^{P}
\label{sy}
\ee

\section{Direct scattering problem:}

In this section we consider the scattering problem of Eq. (\ref{sc}) and try to
address the direct scattering problem, i.e, we construct the eigenfunction 
of the scattering problem of Eq. (\ref{sc}) corresponding to the boundary 
condition that the potentials  ${\bf Q}$ and ${\bf R}$ vanish rapidly as 
$|x|\rightarrow \infty $. The linear dependence 
of different set of eigenfunctions satisfying definite boundary 
conditions at the two spatial infinities will be discussed with the 
construction of the proper eigenvalue, norming constants
and symmetry properties of the eigenfunctions.

\par
In order to find the eigenfunctions, we first construct the asymptotic 
form of wave functions which can be 
obtained by solving the following equation:

\bea
v_{x}&=&
\bp
-ik{\bf I}_{1\times1} & {\bf0}_{1\times 2} \\
{\bf0}_{2\times 1} & ik{\bf I}_{2\times2}  \\
\ep
v
\label{sc1}
\eea
where the matrix in the right hand side is obtained  by imposing the asymptotic limit on the potentials in eq.
(\ref{sc}) and
${\bf 0}_{n \times m}$ denotes $ n \times m$ zero block.. The solutions of the above equation can be readily obtained. We make
the following choices of the asymptotic form of the eigenfunctions:

\bea
\phi(x,k)=
\bp
{\bf I}_{1\times1}\\
{\bf0}_{2\times1}\\
\ep
exp(-ikx),\    
,\   \bar\phi(x,k)=
\bp
{\bf0}_{1\times2}\\
{\bf I}_{2\times2}\\
\ep   
\  exp(ikx)       
\label{ei1}   
\label{b1}
\eea
as $x \rightarrow -\infty$
\bea
\psi(x,k)=
\bp
{\bf0}_{1\times2}\\
{\bf I}_{2\times2}\\
\ep   
\  exp(ikx),\    \bar \psi(x,k)=
\bp
{\bf I}_{1\times1}\\
{\bf0}_{2\times1}\\
\ep
\ exp(-ikx)    
\label{ei2}
\label{b2}      
\eea
as $x \rightarrow +\infty$.\\\\

In order to make the asymptotic form of the eigenfunctions to be 
independent of $x$, we define the following Jost functions:
\bea
M(x,t)=exp(ikx)\phi(x,k),\            \bar{M}(x,k)=exp(-ikx)\bar{\phi}(x,k),\\
N(x,t)=exp(-ikx)\psi(x,k),\            \bar{N}(x,k)=exp(ikx)\bar{\psi}(x,k),
\eea 
where M and $\bar{N}$ are $3 \times 1$ and  N and $\bar{M}
$ are $3\times2$ matrices  
satisfying the constant boundary conditions:

\bea
M(x,k)=
\bp
{\bf I}_{1\times1}\\
{\bf0}_{2\times1}\\
\ep
,\     \bar M(x,k)=
\bp
{\bf0}_{1\times2}\\
{\bf I}_{2\times2}\\
\ep   \nonumber
\eea
as $x \rightarrow -\infty$
\bea
N(x,k)=
\bp
{\bf0}_{1\times2}\\
{\bf I}_{2\times2}\\
\ep  
,\    \bar N(x,k)=
\bp
{\bf I}_{1\times1}\\
{\bf0}_{2\times1}\\
\ep        
\label{bc}   
\eea
as $x \rightarrow +\infty$.\\\\

\noindent With these choices of the Jost functions, The scattering 
problem of Eq. (\ref{sc}) can be transformed into a set 
of integral equations \cite{aab} which can be solved and
up to first order of $k$ the solutions for the Jost functions 
compatible to the boundary conditions (\ref{bc}) may be 
summarized as:

\bea
M(x,k)=
\bp
1-\frac{1}{2ik}\int^{x}_{-\infty}{\bf Q}{\bf R}(x^{'})dx^{'}\\\\
-\frac{1}{2ik}{\bf R}
\ep
+O(k^{-2})
\label{m}
\eea

\bea
\bar{N}(x,k)=
\bp
1+\frac{1}{2ik}\int^{\infty}_{x}{\bf Q}{\bf R}(x^{'})dx^{'}\\\\
-\frac{1}{2ik}{\bf R}
\ep
+O(k^{-2})
\label{n}
\eea

\bea
\bar{M}(x,k)=
\bp
\frac{1}{2ik}{\bf Q}(x,k) \\\\
{\bf I_{2\times2}}+\frac{1}{2ik}\int^{x}_{-\infty}{\bf R}{\bf Q}(x^{'})dx^{'} 
\ep
+O(k^{-2})
\label{m1}
\eea

\bea
N(x,k)=
\bp
\frac{1}{2ik}{\bf Q}(x,k) \\\\
{\bf I_{2\times2}}-\frac{1}{2ik}\int^{\infty}_{x}{\bf R}{\bf Q}(x^{'})dx^{'}
\ep
+O(k^{-2})
\label{n1}
\eea

\noindent If ${\bf Q}, {\bf R} \in L^1({\mathbb R})$, then the functions 
M, N are analytic in the upper half
 whereas $\bar{M}, \bar {N}$ are analytic in the
lower half of the complex k-plane \cite{aab}. It 
should be noted that the expression of the Jost functions in Eqs. 
(\ref{m}-\ref{n1})
have the similar form as in the corresponding local case. However,
the difference will appear through the nonlocal reduction  
${\bf R} =-{\bf Q}^{\bf P}$.

\noindent The above expressions (\ref{m}-\ref{n1}) of the Jost-functions will be 
used together with the results from 
the inverse problem to reconstruct the potentials. 

\subsection{Scattering data: }

If $u_1$ and $u_2$ are two linearly independent solutions of 
the differential equation
$
\frac{dv}{dx}={\bf A}v ,
$
then the Wronskian, i.e $W(u_1,u_2)$, must be non-vanishing.
Since 
$
\frac{dW}{dx}=(tr{\bf A}) W ,
$
we have, using the boundary conditions (\ref{b1}-\ref{b2}) , 
\bea
W(\phi,\bar\phi)=e^{ikx} ,\    \   
W(\psi,\bar\psi)=e^{ikx}
\label{ww1}
\eea
 which indicates
 $\{\phi,\bar\phi\}$ and $\{\psi,\bar\psi\}$ are the two sets of linearly 
independent solutions of the scattering problem.
Since eq. (\ref{2e}) is a 3rd order linear ordinary 
differential equation we can express one
 set of solutions in terms of the linear 
combinations of the other. Thus we can write

\bea
\bp
\phi(x,k)&
\bar{\phi}(x,k)
\ep
=
\bp
\bar{\psi}(x,k)&
\psi(x,k)
\ep
\bp
{\bf a} & {\bf \bar{b}}\\
{\bf b} & {\bf \bar{a}}
\ep
\label{l}
\eea

\noindent where ${\bf{b}}(k)$ is $2 \times 1$, ${\bf{a}}(k)$ is $1\times1$, $\bar{\bf{a}}(k)$ is
 $2\times2$ and $\bar{\bf{b}}(k)$ is $1\times2$ matrix.
Making use of eqs. (\ref{ww1}) and eq. (\ref{l}) 
we get the following result:

\bea
\det
\bp
{\bf a} & {\bf \bar{b}}\\
{\bf b} & {\bf \bar{a}}
\ep
=-1
\eea

The scattering coefficients in terms of the Wronskian of the wave 
functions may be written as:
\bea
W(\phi(x,k),\psi(x,k))=W(\bar{\psi}(x,k),\psi(x,k))\det {\bf a}(k)
\label{w3}\\
W(\bar{\psi}(x,k),\bar{\phi}(x,k))=W(\bar{\psi}(x,k),\psi(x,k))\det {\bar{\bf a}}(k)
\label{w4}
\eea

The proper eigenvalue of the scattering problem (\ref{sc}) is defined to 
be the complex value of k for which it gives a bound state solution that 
decay as $|x|\rightarrow  \infty$.
The asymptotic expressions for the wave functions (\ref{ei1},\ref{ei2})
suggest, for $Im k>0$, $\phi$ decays as $x \rightarrow -\infty$ while 
$\psi$ decays as $x \rightarrow \infty$. Therefore $k_j=\xi_j+i\eta_j$ will 
be a proper eigenvalue in the upper k plane if $\phi$ and $\psi$ are linearly
dependent, thus we must have 
\bea
W(\phi(x,k),\psi(x,k))=0
\eea
and eq. (\ref{w3}) implies that this will happen if $\det{\bf a}=0$.
Thus eigenvalue in the upper k-plane is that value of $k=k_j$ for which
$\det{\bf a}(k_j)=0$. Similarly eq. (\ref{w4}) suggests eigenvalue in the lower k-plane 
is that value of $k=\bar{k_j}$ for which
$\det\bar{{\bf a}}(\bar k_j)=0$. 
None of the eigenfunctions vanishes as $|x|\rightarrow \infty$
if k is real and therefore no proper eigenvalue exists on the real k-axis. 
Further we assume that
$\det{\bf a}\ne 0$, $\det{\bar{\bf a}}\ne0$ on the real k-axis.
When k is a proper eigenvalue in the upper and in the lower k-plane, we may write
\bea
\phi(x,k_j)=\psi(x,k_j){\bf{C}}_j\  \  
\bar{\phi}(x,\bar{k}_j)=\bar{\psi}(x,\bar{k}_j)\bar{\bf{C}}_j,
\eea
respectively,
where ${\bf{C}}_j={\bf{C}}(k_j)$ is a $2\times1$ matrix and $\bar{\bf{C}_j}
=\bar{\bf{C}}(\bar k_j)$ is a $1\times2$  matrix.\\\\

\subsection{Symmetry reduction}

The symmetry reduction ${\bf R}=-{\bf Q}^{P}$ induces a symmetry in 
the wave functions and this in turns induces a symmetry in the scattering data, in 
particular the following symmetry relations holds:

\bea
{M^{up}}^{*}(-x,-k^{*})\equiv N^{dn}(x,k),\    \    {M^{dn}}^{*}(-x,-k^{*})\equiv N^{up}(x,k)\\
\bar{{M}^{up}}^{*}(-x,-k^{*})\equiv \bar{N}^{dn}(x,k),\  
  \    \bar{{M}^{dn}}^{*}(-x,-k^{*})\equiv \bar{N}^{up}(x,k)
\eea

where 

\bea
M^{up}\equiv M_1,\     \ M^{dn}\equiv \bp M_2\\ M_3 \ep \\
N^{up}\equiv \bp N_{11}& N_{12}\ep, \   \ N^{dn}\equiv \bp N_{21} & N_{22}\\ N_{31} & N_{32}\ep
\eea

\bea
\bar{N}^{up}\equiv\bar{N}_1,\     \ \bar{N}^{dn}\equiv\bp \bar{N}_2\\ \bar{N}_3 \ep \\
\bar{M}^{up}\equiv\bp M_{11}& \bar{M}_{12}\ep, 
\   \ \bar{M}^{dn}\equiv\bp \bar{M}_{21} & \bar{M}_{22}\\ \bar{M}_{31} & \bar{M}_{32}\ep
\eea

\noindent This symmetry of the Jost functions is new 
compare to the corresponding local case 
and is solely related to the new nonlocal reduction
${\bf R}=-{\bf Q}^{P}$.

\section{Time evolution:}
In this section we shall consider the other linear 
equation of the Lax pair formulation, i.e, the time 
evolution Eq. (\ref{3e}) to find the time evolution
of the scattering data.
The time evolution Eq. (\ref{3e}) may be written in the following fashion:
\bea
\partial_t v=
\bp
{\bf A} & {\bf B}\\
{\bf C} & {\bf D}
\ep
v
\eea

where

\bea
{\bf A}=2ik^2+i{\bf Q}{\bf R},\    \ {\bf B}=-2k{\bf Q}-i{\bf Q}_{x}\nonumber\\
{\bf C}=-2k{\bf R}-i{\bf R}_{x},\    \ {\bf D}=-2ik^2{\bf I_{2\times2}}-i{\bf R}{\bf Q}.
\eea

Since ${\bf R}, {\bf Q}\rightarrow 0$ as $|x|\rightarrow \infty$,
 time-dependence of $v$ asymptotically 
must satisfy the following equation:

\bea
\partial_t v=
\bp
 A_{\infty} & 0\\
 0 & A_{\infty}{\bf I}_{2\times2}
\ep
v
\eea
 where $A_{\infty}=2ik^2$.
In-order that the solutions of the above equation is compatible
with the boundary conditions imposed in (\ref{b1} -\ref{b2}), we define the following
wave functions:

\bea
\phi^{t}=e^{A_{\infty}t}\phi(x,t),\   \ \bar{\phi^{t}}=e^{A_{\infty}t}\bar{\phi}(x,t)\\
\psi^{t}=e^{A_{\infty}t}\psi(x,t),\   \ \bar{\psi^{t}}=e^{A_{\infty}t}\bar{\psi}(x,t)
\eea

The time evolution of $\phi, \bar{\phi}$, therefore satisfy:

\bea
\partial_t \phi=
\bp
{\bf A}-2ik^2A_{\infty} & {\bf B}\\
{\bf C} & {\bf D}-A_{\infty}{\bf I}_{2\times2}
\ep
\phi
\eea

\bea
\bar{\partial}_t \phi=
\bp
{\bf A}-2ik^2A_{\infty} & {\bf B}\\
{\bf C} & {\bf D}-A_{\infty}{\bf I}_{2\times2}
\ep
\bar{\phi}
\eea

\noindent By evaluating the above equations and exploiting the relations in (\ref{l}) we 
get the following equations:

\bea
\partial_t {\bf a}(k)=0,\  \ \partial_t \bar{{\bf a}}(k)=0\\
\partial_t {\bf b}(k)=-2A_{\infty} {\bf b}(k),\  \ \partial_t \bar{{\bf b}}=
2A_{\infty} \bar{{\bf b}}(k)
\eea 

\noindent which gives the following time dependencies
of the scattering co-efficients:

\bea
{\bf a}(k)={\bf a}(k,0),\   \  \bar{{\bf a}}(k)= \bar{{\bf a}}(k, 0)\\
{\bf b}(k)=e^{-4ik^2t}{\bf b}(k,0),\   \  \bar{{\bf b}}(k)=e^{4ik^2t} \bar{{\bf b}}(k, 0)
\eea
Thus ${\bf a}(k)$ and $\bar{{\bf a}}(k)$ both are constants  indicating 
the eigenvalues, being the zeros of ${\bf a}(k)$ and $\bar{{\bf a}}(k)$ in
the upper and lower half of the k-plane respectively, are also constant.
The time dependence of the norming constants are readily obtained \cite{aab}:
\bea
{\bf C}_j(t)={\bf C}_j(t)e^{-4ik^2t},\  \ \bar{{\bf C}}_j(t)=\bar{{\bf C}}_j(t)e^{4i\bar{k}^2t}
\eea

\section{Inverse scattering problem:}

This section deals with the reconstruction of the potentials by using the RH
approach. Using eq. (\ref{l}) one can obtain, by exploiting the 
analytic properties of the Jost functions, the following
equations governing ${\bf N}(x,k)$ and ${\bf \bar{N}}(x,k)$  \cite{aab}: 
\bea
{\bf N}(x,k)=
\bp
{\bf 0}\\
{\bf I}
\ep
+\sum^{\bar{J}}_{j=1} \frac{e^{-2i\bar{k_j}x}}{(k-\bar
{k_j})}{\bf \bar{N}}_{j}(x){\bf\bar{C}}_j
-\frac{1}{2\pi i}\int^{\infty}_{-\infty}\frac{e^{-2i\xi x}}
{\xi-(k+i0)}{\bf \bar{N}}(x,\xi)\bar{{\bf \rho}}(\xi)d\xi
\label{ee1}
\eea

\bea
\bar{{\bf N}}(x,k)=
\bp
{\bf I}\\
{\bf 0}
\ep
+\sum^{J}_{j=1} \frac{e^{2ik_jx}}{(k-k_j)}{\bf N}_{j}(x){\bf C}_j
+\frac{1}{2\pi i}\int^{\infty}_{-\infty}\frac{e^{2i\xi x}}{\xi-(k-i0)}
{\bf N}(x,\xi){\bf \rho}(\xi)d\xi
\label{ee2}
\eea
where ${\bf N}_{j}(x)={\bf N}(x,k_{j})$, $\bar{{\bf N}}_{j}(x)=\bar{{\bf N}}(x,k_{j})$, 
$\rho={\bf b}{\bf a}^{-1}$ and $\bar{\rho}={\bar{\bf b}}{\bar{\bf a}}^{-1}$
 In order to close the system we evaluate eq. (\ref{ee1}) at $k=k_j$, for $j=1.....J$  and 
eq. ({\ref{ee2}}) at $k=\bar{k_l}$, for $l=1,2....\bar{J}$ and obtain the following results:

\bea
{\bf N}_{j}(x)=
\bp
{\bf 0}\\
{\bf I}
\ep
+\sum^{\bar{J}}_{m=1} \frac{e^{-2i\bar{k}_m x}}{(k_j-\bar{k}_m)}
{\bf \bar{N}}_{j}
(x){\bf\bar{C}}_m
-\frac{1}{2\pi i}\int^{\infty}_{-\infty}\frac{e^{-2i\xi x}}{\xi-k_{j}}{\bf
 \bar{N}}(x,\xi)\bar{{\bf \rho}}(\xi)d\xi
\label{ee3}
\eea

\bea
\bar{{\bf N}_{l}}(x)=
\bp
{\bf I}\\
{\bf 0}
\ep
+\sum^{J}_{j=1} \frac{e^{2ik_jx}}{(\bar{k}_{l}-k_j)}{\bf N}_{j}(x){\bf C}_j
+\frac{1}{2\pi i}\int^{\infty}_{-\infty}\frac{e^{2i\xi x}}{\xi-\bar{k}_{l}}
{\bf N}(x,\xi){\bf \rho}(\xi)d\xi
\label{ee4}
\eea

\noindent The above two equations yield a linear algebraic-integral system 
of equations that, in principle,
solve the inverse problem for ${\bf N}(x,k)$ and $\bar{{\bf N}}(x,k)$.

\section{One soliton solution:}

In this section we shall use the results we got
from direct and inverse scattering and find the one 
soliton solution for the two component vector NLSE. 
In case when the scattering problem does not possess
any eigenvalue from the real k-axis and $\rho(k)=0$,
$\bar{\rho(k)}=0$ for all k, we have, using eqs. (\ref{ee3},\ref{ee4})
\bea
{\bf N}_{j}(x)=
\bp
{\bf 0}\\
{\bf I}
\ep
+\sum^{\bar{J}}_{m=1} \frac{e^{-2i\bar{k_m}x}}{(k_j-\bar{k_m})}
{\bf \bar{N}}(x,k_{j}){\bf\bar{C}}_m
\label{e5}
\eea

\bea
\bar{{\bf N}_{l}}(x)=
\bp
{\bf I}\\
{\bf 0}
\ep
+\sum^{J}_{j=1} \frac{e^{2ik_jx}}{(\bar{k}_{l}-k_j)}{\bf N}_{j}(x){\bf C}_j
\label{e6}
\eea 

\noindent To obtain the one soliton solution, we take $J=\bar{J}=1$. It is interesting to 
note that in the corresponding local case the symmetry reduction gives a
relation between the two wave functions analytic in the upper half of the k-plane, on the
other hand the symmetry reduction in the nonlocal case relates 
 the wave functions, analytic in the upper half of the k-plane to that of
analytic in the lower half of the k-plane. This fact manifest itself in the fact that
in the nonlocal case $k$ and $\bar{k}$ are independent in contrast to the
  local case where we have $\bar{k}=k^*$. 
 If we now write the above two equations in the component form then we 
get (hereafter we shall suppress the indices j or l ):

\bea
N_{11}=\frac{e^{-2i\bar{k}x}\bar{N}_1\bar{C}_1}{k-\bar{k}},\      \ N_{12}=
\frac{e^{-2i\bar{k}x}\bar{N}_1\bar{C}_2}{k-\bar{k}}\nonumber\\
N_{21}=1+\frac{e^{-2i\bar{k}x}\bar{N}_2\bar{C}_1}{k-\bar{k}},\       \ N_{22}=
\frac{e^{-2i\bar{k}x}\bar{N}_2\bar{C}_2}{k-\bar{k}}\nonumber\\
 N_{31}=\frac{e^{-2i\bar{k}x}\bar{N}_3\bar{C}_1}{k-\bar{k}},\        \ N_{32}
=1+\frac{e^{-2i\bar{k}x}\bar{N}_3\bar{C}_2}{k-\bar{k}}
\label{nn}
\eea
and also,

\bea
\bar{N}_1=1+\frac{e^{2ikx}(N_{11}C_1+N_{12}C_2)}{\bar{k}-k}\nonumber\\
\bar{N}_2=\frac{e^{2ikx}(N_{21}C_1+N_{22}C_2)}{\bar{k}-k}\nonumber\\
\bar{N}_3=\frac{e^{2ikx}(N_{31}C_1+N_{32}C_2)}{\bar{k}-k}
\label{nnn}
\eea

\noindent Comparing eqs. (\ref{nn}-\ref{nnn}) and eqs. (\ref{n}-\ref{n1}) for large k we get
\bea
r_1(x)=-2i e^{2ikx} (N_{21}C_1+N_{22}C_2)\\
r_2(x)=-2i e^{2ikx} (N_{31}C_1+N_{32}C_2)
\eea
and
\bea
q_1(x)=2i e^{-2i\bar{k}x} \bar{N}_1\bar{C}_1\\
q_2(x)=2i e^{-2i\bar{k}x} \bar{N}_1\bar{C}_2.
\eea

\noindent Solving for the N's and $\bar{N}$'s, using equations (\ref{nn},\ref{nnn}), we get

\bea
r_1(x)=-2i\frac{C_1e^{2ikx}}{1+\frac{e^{2ix(k-
\bar{k})(C_1\bar{C}_1+C_2\bar{C}_2)}}{(k-\bar{k})^2}}\nonumber\\
r_2(x)=-2i\frac{C_2 e^{2ikx}}{1+\frac{e^{2ix(k-
\bar{k})(C_1\bar{C}_1+C_2\bar{C}_2)}}{(k-\bar{k})^2}}
\label{r}
\eea
and
\bea
q_1(x)=2i\frac{\bar{C}_1e^{-2i\bar{k}x}}{1+\frac{e^{2ix(k-\bar{k})
(C_1\bar{C}_1+C_2\bar{C}_2)}}{(k-\bar{k})^2}}\nonumber\\
q_2(x)=2i\frac{\bar{C}_2e^{-2i\bar{k}x}}{1+\frac{e^{2ix(k-\bar{k})
(C_1\bar{C}_1+C_2\bar{C}_2)}}{(k-\bar{k})^2}}
\label{q}
\eea

\noindent with $k=i\eta$ and $\bar{k}=-i\bar{\eta}$, $\eta \ne \bar{\eta}$, takes
care of the fact that $k$ and $\bar{k}$ are independent in general,
 and $\bar{\eta},\eta >0$. 

\noindent The symmetry reduction (\ref{sy}) induces the following restrictions on the possible
forms of the norming constants:
\bea
C_1(0)=\frac{1}{\sqrt{2}}(\eta+\bar{\eta})e^{i(\theta_1+\frac{\pi}{2})},\  \  \bar{C_1}(0)=
\frac{1}{\sqrt{2}}(\eta+\bar{\eta})e^{i(\bar{\theta}_1+\frac{\pi}{2})},\nonumber\\
C_2(0)=\frac{1}{\sqrt{2}}(\eta+\bar{\eta})e^{i(\theta_2+\frac{\pi}{2})},\  \  \bar{C_2}(0)=
\frac{1}{\sqrt{2}}(\eta+\bar{\eta})e^{i(\bar{\theta}_2+\frac{\pi}{2})},
\eea
and
\bea
(\theta_1+\bar{\theta}_1)=(\theta_2+\bar{\theta}_2),
\eea
with this choices and considering the time dependence of the norming constants,
  eq. (\ref{r}-\ref{q}) reduce to the following form:

\bea
q_1(x)=-\sqrt{2}\frac{(\eta+\bar{\eta})e^{i\bar{\theta}_1}e^{-4i\bar{\eta}^{2}t}
e^{-2\bar{\eta}x}}{1+e^{i(\theta_1+\bar{\theta}_1)}e^{-4i(\bar{\eta}^{2}-
\eta^{2})t}e^{-2x(\eta+\bar{\eta})}}\nonumber\\
q_2(x)=-\sqrt{2}\frac{(\eta+\bar{\eta})e^{i\bar{\theta}_2}
e^{-4i\bar{\eta}^{2}t}e^{-2\bar{\eta}x}}{1+e^{i(\theta_1+\bar{\theta}_1)}
e^{-4i(\bar{\eta}^{2}-\eta^{2})t}e^{-2x(\eta+\bar{\eta})}}\nonumber\\
\eea

\bea
r_1(x)=\sqrt{2}\frac{(\eta+\bar{\eta})e^{i\theta_1}e^{4i\eta^{2}t}
e^{-2\eta x}}{1+e^{i(\theta_1+\bar{\theta}_1)}
e^{-4i(\bar{\eta}^{2}-\eta^{2})t}e^{-2x(\eta+\bar{\eta})}}\nonumber\\
r_2(x)=\sqrt{2}\frac{(\eta+\bar{\eta})e^{i\theta_2}e^{4i\eta^{2}t}
e^{-2\eta x}}{1+e^{i(\theta_1+\bar{\theta}_1)}
e^{-4i(\bar{\eta}^{2}-\eta^{2})t}e^{-2x(\eta+\bar{\eta})}}
\eea

\noindent In the special case, when $\eta=\bar{\eta}$, ${\theta_1=-\bar{\theta}_1}
$ and ${\theta_2=-\bar{\theta}_2}$ (as in the case of local VNLSE), 
we get:

\bea
q_{1}(x,t)=-\sqrt{2}\eta e^{-i\theta_1}e^{-4i{\eta}^{2}t} sech(2\eta x)\nonumber\\
q_{2}(x,t)=-\sqrt{2}\eta e^{-i\theta_2}e^{-4i{\eta}^{2}t} sech(2\eta x)
\eea

\noindent which is the solution of the two component local VNLSE of the form of Eq. (\ref{1e})
with the nonlocal nonlinear interaction term being replaced by a local one.

\section{Conserved quantities:}

We have shown ${\bf a}(k)$ is independent of time. Further the integral expression of  ${\bf a}(k)$
may be written as \cite{aab}:

\bea
{\bf a}(k)={\bf I}_N+\int^{\infty}_{-\infty}{\bf Q}(x){\bf M}^{(dn)}(x,k)dx.
\eea
Since ${\bf a}(k)$ and therefore ${\bf a}(k)-{\bf I}_N$ is analytic in the upper 
k-plane with ${\bf a}(k)-{\bf I}_N \rightarrow 0$ as $|k|\rightarrow 0$ we may 
expand ${\bf a}(k)-{\bf I}_N$ in the Laurent series expansion with each of the 
coefficients being a constant.
Since the number of coefficients in the Laurent series is infinity, 
we get an infinite number of conserved quantities in this case.
 In general the conserved quantities may be written 
as:

\bea
\nu_n=\int^{\infty}_{-\infty}{\bf Q}(x){\bf W}_n(x)dx
\eea
with ${\bf W}_n(x)$  given as
\bea
{\bf W}_n=-\frac{d^n}{dx^n}{\bf R}(x)-\sum^{n}_{m=1} \frac{d^{n-m}}{dx^{n-m}}\left({\bf R}(x)\int^{x}_{-\infty}{\bf Q}(x){\bf W}_{m-1}\right)
\eea

The first few constant quantities may be 
written as:
\bea
\nu_0&=&-\int^{\infty}_{-\infty}{\bf Q}(x){\bf R}(x)dx
\label{nu1}\\
\nu_1&=&-\int^{\infty}_{-\infty}{\bf Q}(x){\bf R}_x(x)dx +\int^{\infty}_{-\infty}dx{\bf Q}(x){\bf R}(x)\nonumber\\
&.&[\int^{x}_{-\infty}{\bf Q}(x^{'}){\bf R}(x^{'})dx^{'}]
\label{nu2}\\
\eea

\noindent We now show, with the reduction ${\bf R}=-{\bf Q}^P$, the conserved quantity 
\bea
\nu_0 &=&\int^{\infty}_{-\infty}\left( \rho_1+\rho_2\right) dx, \    \rho_j(x,t)=(q_j(x,t)q^*_j(-x,t), j=1,2 
\eea
is
real valued. We decompose the fields
$q_1$ and $q_2$ as a sum of parity-even and parity-odd terms \cite{dp}:
\bea
q_j( x, t)=q_{je}(x, t)+q_{jo}(x, t),\\
\label{sa}
\eea
where 
\bea
q_{je}(x, t)=\frac{q_j(x,t)+q_j(-x,t)}{2}, \ \
q_{jo}(x, t)=\frac{q_j(x,t)-q_j(-x,t)}{2}.\\
\eea
\noindent  This decomposition of the fields allows to express the densities
$\rho_j$ 
as sum of a real-valued parity-even term and a parity-odd term which is purely
imaginary. In particular,
\bea
\rho_j(x, t)=\rho_{jr}(x,t) + \rho_{jc}(x,t)\\
\label{AB}
\eea
with
\bea
\rho_{jr}(x, t)={\mid q_{je}(x, t) \mid}^2 -
{\mid q_{jo}(x, t) \mid}^2, \ \
\rho_{jc}(x, t)=q^{*}_{je}(x, t)q_{jo}(x, t)-q^{*}_{jo}(x, t)q_{je}( x, t),\\
\label{exAB}
\eea

\noindent 
The densities,  $\rho_{jr}(x,t)$, $\rho_{jc}(x,t)$ 
possess the following properties:
\bea
&& \rho_{jr}^{*}(x,t)=\rho_{jr}(x,t),
\ \  \rho_{jr}(- x,t)=\rho_{jr}(x,t),\nonumber \\
&& \rho_{jc}^{*}(x,t)= -\rho_{jc}(x,t),
\ \  \rho_{jc}(-x,t)=-\rho_{jc}(x,t), \
\eea
\noindent which indicates that the densities are  complex-valued functions. However, 
$\nu_0$, as defined in terms of eq. (\ref{nu1}), does not receive any contribution from
the parity-odd purely imaginary term $\rho_c(x,t)$ since this part
vanishes during the integration over the densities. Thus in spite of 
non-Hermiticity we recover the reality
of the conserved quantity $\nu_0$. Similar decomposition of the fields shows that $N^1$ and $N^2$ 
are complex in general but $N^3$ is real-valued. Using similar calculations as in Ref. \cite{dp},
it can be shown that the Hamiltonian is also real valued.

\section{Inhomogeneous nonlocal VNLSE with space-time modulated nonlinear interaction term:}
In this section we consider an inhomogeneous version of the two 
component non-local VNLSE with 
space-time modulated nonlinear interaction term.
  This kind of equations appear in the 
study of multi component BEC. There exist methods for finding exact solutions of this kind equations as discussed
in Ref \cite{1jbb}\cite{jbb}\cite{V}. Here we have used a mapping  to the standard VNLSE  through similarity 
transformation to obtain exact solution
of the inhomogeneous two 
component VNLSE  with space-time modulated 
nonlinear interaction term of the following form:
\bea
i\frac{\partial {\bar q}_j}{\partial t}=-\frac{1}{2}\frac{\partial^2{\bar q}_j}{\partial x^2}+(V_j+i W_j){\bar q}_j+\sum^{2}_{k=1}(g_{jk}(x,t){\bar q}_k(x,t){\bar q}_k^*(-x,t)){\bar q}_j,\     j=1,2\nonumber\\
\label{si}
\eea

\noindent Here $ V_j(x,t)+i W_j(x,t)$  is chosen to be the external potential with $ V_j(x,t)$ and $ W_j(x,t)$ are 
being
 the real and imaginary parts respectively and $g_{ij}(x,t)$ with $i, j=1, 2$ is the space-time
modulated nonlinear interaction term. It is interesting to note that
the external potential becomes ${\cal PT}$ symmetric when $ V_j(-x,-t)= V_j(x,t)$
and $W_j(-x,-t)=- W_j(x,t)$. 

We now use the similarity transformation 
\bea
{\bar q}_j=\rho(x,t) \exp(i\phi_j(x,t))q_j
\label{sitr}
\eea
to reduce Eq. (\ref{si}) to the following Eq.

\bea
\mu_jq_j=-\frac{1}{2}\frac{\partial^2q_j}{\partial x^2}+\sum^2_{k=1}(G_{jk}q_k(x,t)q^*_k(-x,t))q_j
\label{si2}
\eea
where $G_{jk}$ is constant. 
Thus the known exact solutions of Eq. (\ref{si2})\cite{akas} can be used to 
generate a large class of solvable inhomogeneous  nonlocal
VNLSE with space-time modulated nonlinear interaction term
 of the type of (\ref{si}). 

We  find  that eq.(\ref{si}) can be  reduced to the stationary non-local  VNLSE of Eq. (\ref{si2}) only when
$X(x, t)$ is an odd function of $x$, i.e.,
\be
 X(-x, t)=-X(x, t), 
\label{TrX}
\ee
and  in addition the following consistency conditions hold simultaneously:
\bea
2\rho \rho_{t} +(\rho^2\phi_{jx})_x&=&2\rho^2  W_j(x,t)
\label{e1}\\
(\rho^2 X_x)_x&=&0
\label{e2}\\
X_t +\phi_{jx} X_x &=&0
\label{e3}\\
 V_j(x,t)&=&\frac{1}{2}{\rho}^{-1}\rho_{xx}-\phi_{jt}-\frac{1}{2}\phi^2_{jx}-\mu_j X^2_x
\label{e4}\\
{\bar g}_{jk}(x,t)&=&\frac{G_{jk}X^2_x}{\rho(x,t)\rho(-x,t)exp(\phi_j(x,t)-\phi_j(-x,t))}
\label{e5}
\eea

\noindent Thus the similarity transformation gives rise to the same set of equations
as in the corresponding scalar case discussed in Ref. \cite{dp}.
  Similar arguments as in Ref. \cite{dp} suggest that $\rho$ and $\phi$
are even functions of $x$. Thus we may now write ${\bar g}_{jk}(x,t)$ as:

\bea
{\bar g}_{jk}(x,t)&=&\frac{G_{jk}X^2_x}{\rho^2(x,t)}
\label{e6}
\eea

Also solving  Eqs. (\ref{e2}-\ref{e3}) we get:

\bea
\rho&=&\sqrt{\frac{\delta(t)}{X_x}}\nonumber\\
\phi_j&=&-\int\frac{X_t}{X_x}dx+\phi_{j0}(t)
\label{enew}
\eea

where $\delta$ and $\phi_{j0}$ are some functions of time. 
\noindent We now discuss some of the examples to illustrate the results.\\\\

\section{Inhomogeneous autonomous non-local VNLSE}

We consider a special type of similarity transformation in which case

\bea
\rho= \rho(x), \    \phi_j(x,t)=-E_jt, \    X=X(x).
\eea

This choice gives $W_j=0$ and Eq. (\ref{e3}) is satisfied automatically. Eqs.
(\ref{e2},\ref{e4},\ref{e5}) take the following forms respectively

\bea
X(x)&=&\int_0^x \frac{ds}{\rho^2}
\label{e7}\\
 V_j(x)&=&\frac{\rho_{xx}}{2\rho}+\{E_j-\frac{\mu_j}{\rho^4}\}
\label{e8}\\
{\bar g}_{jk}(x)&=&\frac{G_{jk}}{\rho^6(x)}
\label{e9}
\eea

From Eqs (\ref{e8}) and (\ref{e9}), it immediately follows 
that $V_j$ and ${\bar{\bf G}}$ are also even functions of $x$. Thus
the similarity transformation technique is valid for non-local two component
VNLSE only when the confining potential and the nonlinear interaction terms
are even functions of spatial coordinate.

Eq. (\ref{e8}) reduces to the Ermakov-Pinney equation of the 
following form \cite{1jbb}:

\bea
\frac{1}{2} \rho_{xx}+ \left [ E_j-V_j(x) \right ] \rho = \frac{\mu_j}{\rho^3}
\label{ep}
\eea
 with the solution of the form:

\be
\rho=\left[a \phi^{2}_1(x) + 2 b \phi_1(x) \phi_2(x) +
c \phi_2^2(x) \right ]^{\frac{1}{2}},
\label{so}
\ee
where a, b, c  are constants and $\phi_1(x)$, $\phi_2(x)$ are the two
linearly independent solutions of the time independent Schrödinger equation,
\be
-\frac{1}{2} \phi_{xx}+V_j(x)\phi(x)=E\phi(x).
\label{s}
\ee
and the constant $\mu$ is determined as,
$\mu= (ac-b^2) \left [ \phi_1^{\prime}(x) \phi_2(x) -\phi_1(x)
\phi_2^{\prime}(x) \right ]^2 $.

\noindent Since $V_j$ is an even function of $x$, the requirement of definite 
parity of $\rho$ can always be ensured by suitable choosing the constants
a, b, c.

\subsection{Example}

\begin{enumerate}
\item {\bf Vanishing External Potentials}\\

In this case we choose $V_1=V_2=0$. Solving  Eq. (\ref{e8}) for this
particular choice of $V_j$ and putting the solution in Eqs. (\ref{e8}-\ref{e9}) we have, for $E>0$

\be
\rho(x) = \left [ 1 + \alpha \cos (\omega x) \right ]^{\frac{1}{2}}, \
{\bar g}_{jk}(x,t) =  G_{jk} \left [ 1 + \alpha \cos (\omega x) \right ]^{-3},\  
\label{rhog}
\ee
where $\omega= 2 \sqrt{2 {\mid E \mid}}$ and $\mu=(1-\alpha^2) E$. The
transformed co-ordinate $X(x)$ is determined as,
\bea
X_+(x) & = & \frac{2}{\omega \sqrt{1-\alpha^2}} \tan^{-1} \left [
\sqrt{\frac{1-\alpha}{1+\alpha}} \tan(\frac{\omega x}{2} )\right ] \ for \
{\mid \alpha \mid} < 1,\nonumber \\
X_-(x)&=& \frac{1}{\omega \sqrt{\alpha^2-1}} \ln \left [\frac{ 
\tan(\frac{\omega x}{2}) + \sqrt{\frac{\alpha+1}{\alpha-1}}}
{ \tan(\frac{\omega x}{2}) - \sqrt{\frac{\alpha+1}{\alpha-1}}}
 \right ] \ for \ {\mid \alpha \mid} > 1,
\eea
where the subscripts refer to the fact that $\mu$ is positive for the solution
$X_+(x)$, whereas it is negative for $X_-(x)$.

The solution of Eq. (\ref{si}) may now be written using Eq. (\ref{sitr}) as:
\bea
{\bar q}_j(x)= \rho(x,t)\exp(-iE_jt)q_j,
\label{sitr1}
\eea
 with $\rho$ is given by Eq. (\ref{rhog}) 
and $q_j$ is any solution of Eq. (\ref{si2}).\\

For $E<0$, we have
\be
\rho(x)= \cosh^{\frac{1}{2}} (\omega x), \ \
g_{jk}(x) =  G_{jk} \cosh^{-3}(\omega x), \  \ \ \
X(x)= - \frac{1}{\omega} \cos^{-1} (\tanh (\omega x))
\ee
where $\mu$ is determined as $\mu= 2 {\mid E \mid}$ which is positive-definite.

\item {\bf Harmonic Confinement}\\

In this case we take  $V_1(x)=V_2=\frac{1}{2} x^2$ and $E=0$ and (\ref{e7}-\ref{e9}) is solved to obtain:
\be
\rho(x)=e^{\frac{x^2}{2}}, \
g_{jk}(x)=  G_{jk} e^{-3 x^2}, \  \ \ \ 
X(x)= \frac{\sqrt{\pi}}{2} erf x.
\ee
with $\mu=0$ and $-\frac{\sqrt{\pi}}{2} \leq X \leq \frac{\sqrt{\pi}}{2}$.

\item {\bf Distinct $V_1$ and $V_2$}\\
\end{enumerate}
We make the choice $V_1=0$ and $E_1=E_2$. This yields
$\rho(x) = \left [ 1 + \alpha \cos (\omega x) \right ]^{\frac{1}{2}}$ for $E_1>0$ as before, also from Eq. (\ref{e8})
we get 
\bea
V_2=\frac{(\mu_1-\mu_2)}{(1+\alpha \cos(\omega x))^2}.
\eea

\section{Non-autonomous nonlocal VNLSE:}

 The condition  (\ref{TrX}) on $X(x,t)$ may be implemented in various ways. Here we consider 
two cases one is separable and other is non-separable in terms of the 
arguments of $X(x,t)$. These choices will make the coefficients of nonlocal VNLSE time dependent.

\subsection {Non-separable case:}

We consider the following ansatz,
\be
X(t,x)=F(\xi), \ \xi(t,x) \equiv \gamma(t) x, \ F(-\xi) = -F(\xi),
\label{oddcondi}
\ee
where $\gamma(t)$ is some function of time. In this case the 
condition (\ref{TrX}) prevents to add any purely time dependent term
in the expression of $X(x)$ which is the particularity of the nonlocal case.
In this case $W_j=0$ and Eqs. (\ref{enew}), (\ref{e6}) and (\ref{e4}) yield:
\bea
\phi_j(x,t) & = & -\frac{\gamma_t}{2 \gamma} x^2 + \phi_{j0}(t), \ \
\rho(x,t) = \sqrt{\frac{\gamma}{F^{\prime}(\xi)}},\ \
g_{jk}(x,t) = { G_{jk} \gamma} \left ( F^{\prime}(\xi) \right )^{3}
,\nonumber \\
V_j(x,t) & = & \frac{\gamma^2}{8 (F^{\prime}(\xi))^2} \left [
3 \left \{ F^{\prime \prime}(\xi) \right \}^2 - 2 F^{\prime \prime}(\xi)
 F^{\prime \prime \prime}(\xi) - 8 \mu_j \left \{ F^{\prime}(\xi)
\right \}^4 \right ] + \frac{1}{2} \omega(t) x^2 - \phi_{j0t}
\eea
with $\delta=\gamma^2$ and $\omega(t)$
 is determined from the equation:
\be
u_{tt} + \omega(t) u=0.
\label{mathieu}
\ee
where $u=\gamma^{-1}$ .

\noindent Interestingly, the above ansatz gives rise to harmonic confinement
irrespective of the choice of $F(\xi)$.

\subsubsection{Separable case}
In this case the expression for $X(x,t)$ is chosen to be separable in terms
of its arguments with the spatial part is taken to be odd with respect to $x$ in order
to respect the condition (\ref{TrX}). In particular, we choose 
\be
X(x,t) \equiv \alpha(t)f(x), \ \ f(-x)= - f(x).
\ee
With this choice of $X$, eqs. (\ref{enew}), (\ref{e6}), (\ref{e1}) and (\ref{e4}) respectively reduce to the following form: 
\begin{eqnarray}
\rho(x,t) &=&\sqrt{\frac{\delta(t)}{\alpha(t)f'(x)}}
\label{exrho}, \ \
\phi(x,t) = -\frac{\alpha_t}{\alpha(t)} \int dx \frac{f(x)}{f'(x)}+\phi_{j0}, \
g_{jk}(x,t) = \frac{ G_{jk} \alpha^{3}}{\delta} {(f^{\prime})}^{3}
\label{exphi}\\
W_j(x, t)&=&\frac{1}{2\alpha(t)\delta(t)}(\delta_t\alpha-2 \alpha_t\delta)
+ \frac{\alpha_t}{\alpha}(\frac{f'' f}{f'^2}),
\label{exW}\\
V_j(x, t)&=&-(\frac{2f'''f'-3f''^2}{8f'^2})+\frac{\alpha_{tt}\alpha-
\alpha^2_t}{\alpha^2}\int \frac{f(x)}{f'(x)}dx-
\frac{\alpha^2_tf^2}{2\alpha^2f'^2}-\mu_j\alpha^2f'^2-\phi_{jot},
\label{exV}
\end{eqnarray}
\noindent  where$f^{\prime}(x) =\frac{df}{dx}$. In this case $W_1=W_2$. 
Since $\phi_0(t)$ is a purely time dependent term, its contribution 
can always be removed from $V(x,t)$ through a phase rotation. It is interesting to
note that whenever $\alpha$ and $\delta$ have definite parity, $V(x,t)$ and 
$W(x,t)$ become even and odd under ${\cal PT}$ respectively. This also implies
that the external potential $v=V+iW$ becomes ${\cal PT}$ symmetric whenever  
$\alpha$ and $\delta$ have definite parity.

\section{Summary and discussions}

In this paper we have considered a two component nonlocal  
VNLSE, with the self-induced potential being ${\cal PT}$ symmetric. We
construct a Lax pair formulation for this equation and employ inverse 
scattering transformation  to obtain the soliton solution for the
 two component
nonlocal vector NLSE. This system is found to possess an infinite
number of conserved quantities. The expression for some of the conserved charges 
are written explicitly. The conserved charges are in general non-Hermitian as in the corresponding one component case.
 However, by decomposing the fields into a
parity even and  a parity odd terms, It is shown that
some of the conserved
charges are real-valued. Further, an inhomogeneous  
version of nonlocal VNLSE with space-time modulated nonlinear interaction term
 is considered and a mapping of this equation to the normal VNLSE through similarity transformation
 is employed to construct the solutions. 

\section{Acknowledgement}
This work is supported by a grant({\bf DST Ref. NO.: SR/S2/HEP-24/2012})
from Science \& Engineering Research Board(SERB), Department of Science
\& Technology(DST), Govt. of India. {\bf DS} acknowledges a research
fellowship from DST under the same project.

\end{document}